# Ferromagnetically correlated clusters in semi-metallic Ru$_2$NbAl Heusler alloy

Sanchayita Mondal[1,2], Chandan Mazumdar[2], R. Ranganathan[2]

[1]*Maharaja Manindra Chandra College, 20 Ramkanto Bose Street, Kolkata 700003,*
*West Bengal, India and*

[2]*Condensed Matter Physics Division, Saha Institute of Nuclear Physics,*
*1/AF, Bidhannagar, Kolkata 700064, India*

Eric Alleno[3]

[3]*Université Paris-Est, Institut de Chimie et des Matériaux Paris-Est,*
*UMR 7182 CNRS UPEC, 2 rue H. Dunant, 94320 THIAIS, France*

P.C. Sreeparvathy[4], V. Kanchana[4]

[4]*Department of Physics, Indian Institute of Technology Hyderabad, Kandi, Medak*
*502285, Telengana, India*

G. Vaitheeswaran[5]

[5]*Advanced Centre of Research in High energy Materials (ACRHEM),*
*University of Hyderabad,*
*Prof. C. R. Rao Road, Gachibowli, Hyderabad 500046, Telengana, India*

(Dated: July 17, 2018)

## Abstract

In this work, we report the structural, magnetic and electrical and thermal transport properties of the Heusler-type alloy Ru$_2$NbAl. From the detailed analysis of magnetization data, we infer the presence of superparamagnetically interacting clusters with a Pauli paramagnetic background, while short-range ferromagnetic interaction is developed among the clusters below 5 K. The presence of this ferromagnetic interaction is confirmed through heat capacity measurements. The relatively small value of electronic contribution to specific heat, $\gamma$ ($\sim 2.7$ mJ/mol-K$^2$), as well as the linear nature of temperature dependence of Seebeck coefficient indicate a semi-metallic ground state with a pseudo-gap that is also supported by our electronic structure calculations. The activated nature of resistivity is reflected in the observed negative temperature coefficient and has its origin in the charge carrier localization due to antisite defects, inferred from magnetic measurements as well as structural analysis. Although the absolute value of thermoelectric figure of merit is rather low ($ZT = 5.2 \times 10^{-3}$) in Ru$_2$NbAl, it is the largest among all the reported non-doped full Heusler alloys.

## I. INTRODUCTION

Since their discovery in 1903, Heusler alloys, having a general formula $X_2YZ$ ($X/Y$ are transition metals, $Z = p$-block elements), have constantly drawn the attention of the researchers due to their remarkable magnetic and transport properties [1–3]. The Heusler alloys crystallize in the $L2_1$ structure (space group: $Fm\bar{3}m$) which consists of four interpenetrating *fcc* sublattices where the $X$-atoms are located at $(\frac{1}{4}\frac{1}{4}\frac{1}{4})$ and $(\frac{3}{4}\frac{3}{4}\frac{3}{4})$ positions, whereas the $Y$ and $Z$ atoms are at $(\frac{1}{2}\frac{1}{2}\frac{1}{2})$ and $(0\,0\,0)$ positions, respectively [3, 4]. This family of compounds includes weak ferromagnets, antiferromagnets, ferrimagnets, half-metallic ferromagnets, metals, semi-metals as well as semiconductors [3]. Valence electron count (VEC) per formula unit of these alloys has a great influence in determining the wide variations in physical properties. For example, the total magnetic moment (M) per unit cell depends on the VEC and is generally estimated to be $M(\mu_B) = |VEC-24|$ (Slater-Pauling rule) [5]. Many members of this class established the validity of the Slater-Pauling rule, *e.g.*, $Mn_2VAl$ (VEC 22) has a moment of 1.94 $\mu_B$/f.u. at 5 K [6] whereas $Co_2FeSi$ (VEC 30) exhibits a moment of 5.97 $\mu_B$/f.u. at 5 K [7]. Thus, Heusler alloys with VEC 24 are expected to be nonmagnetic with a vanishing total magnetic moment per unit cell. Several materials having VEC 24 that have been discovered till now, *e.g.*, $Fe_2VAl$, $Fe_2VGa$, $Fe_2TiSn$ *etc.* [8–10], are indeed found to be nonmagnetic. Interestingly, in all these compounds having VEC 24, a narrow gap or pseudo-gap has been found in the vicinity of the Fermi level and hence they exhibit semiconducting or semimetallic behaviour [8, 10, 11]. However, various inconsistencies are also reported in the experimental measurements of magnetic properties of these alloys. As for example, in spite of having VEC 24, cluster glass behaviour has been reported in $Fe_2VAl$ [12]. Such discrepancies arise primarily due to the presence of antisite defects and disorders introduced during the synthesis and annealing process [13–16]. In the case where the atoms on the $(0\,0\,0)$ and $(\frac{1}{2}\frac{1}{2}\frac{1}{2})$ positions are fully exchanged, the Heusler alloy then adopts the averaged B2 structure type whereas in the extreme case where all the sites are randomly occupied by all the atoms, the alloys adopt the A2 structure type [3]. Several kinds of antisite defects have been reported at low concentration in Heusler alloys: in $Fe_2VAl$, for instance, $Fe_V$ and $Fe_{Al}$ are experimentally known to occur [16, 17]. According to a theoretical study [18], $V_{Al}$ antisite defects influence neither the magnetic properties nor the transport properties whereas the $Fe_V$ and $Fe_{Al}$ defects bear a magnetic moment ($\sim 4$ $\mu_B$) and give rise to electronic states in the pseudo-gap. Having a narrow gap/pseudo-gap in the vicinity of the Fermi level, these compounds are considered to be suitable for thermoelectric applications. The efficiency of a thermoelectric device varies like the Carnot efficiency and the dimensionless figure-of-merit of its constituting materials, $ZT = \frac{S^2 T}{\rho \kappa}$, where $S$ is the Seebeck coefficient, $\rho$ and $\kappa$ are the electrical resistivity and thermal conductivity, respectively. For a thermoelectric material, a value of $ZT$ close to or larger than 1 is generally considered to be large enough to give rise to applications. The highest value of $ZT$ at room temperature (RT) has been found in the compound $Bi_2Te_3$ ($ZT \sim 1$) [19]. However, Te is toxic and expensive, which makes $Bi_2Te_3$ commercialized only in niche markets like localized or silent cooling. Pristine Heusler alloys display $ZT \sim 10^{-3}$ at RT whereas careful doping, leads for instance, to $ZT \sim 0.2$ in $Fe_2V_{0.9}W_{0.1}Al$ [20]. Heusler alloys with VEC 24 are also considered for applications such as the magnetic information storage where they could play the role of a thin non-magnetic buffer layer sandwiched between two ferromagnetic thin layers of Heusler alloy acting as a spin polarizer in a recording head [21, 22].

In this work, we report a detailed study of the structural, magnetic and transport properties of the Heusler alloy $Ru_2NbAl$, whose crystal structure have only been reported so far [23]. Band structure calculations have also been performed for better understanding of the ground state properties. Our experimental data and calculations suggest that this compound is a semi-metal. We have also measured $ZT = 5.2 \times 10^{-3}$ as a value of the thermoelectric figure of merit in $Ru_2NbAl$ at RT. This value, though relatively small, is found to be one of the largest among the non-doped Heusler alloys reported in the literature.

## II. EXPERIMENTAL AND COMPUTATIONAL METHODS

$Ru_2NbAl$ was prepared by a melting process taking stoichiometric amounts of the constituent elements Ru (>99.9%), Nb (>99.9%) and Al (>99.9%) in an arc furnace on a water cooled copper hearth under a flowing Ar atmosphere. The resultant ingot was melted several times, flipped after each melting, to promote homogeneity. The weight loss during the whole process was found to be less than 0.5%. Following the post-synthesis sample treatment procedure reported earlier in $Fe_2VAl$ [24], we have also annealed the as-cast ingot of $Ru_2NbAl$ at 1273 K for 48 hours in a vacuum sealed quartz tube and then quenched in ice-water. After cleaning the surface of the sample, it was again annealed at 1223 K for 12 hours following the same procedure. The sample was then cut in appropriate shapes and polished. The appropriately shaped sample was again annealed for 2 hours at 1173 K using the similar procedure to remove any surface strain that could has developed due to the mechanical stress in the process of cutting and polishing, as some Heusler alloys are indeed found to be highly prone to coldwork [24, 25]. The elemental composition of the annealed material had been estimated by the wavelength dispersive spectroscopy based Electron Probe Micro-Analysis (EPMA) technique [Model: SX 100, Cameca, France]. An essentially single phase nature of $Ru_2NbAl$ was identified by powder X-ray diffraction (XRD) technique at room temperature using Cu $K_\alpha$ radiation in a powder diffractometer having a rotating anode X-ray source at 9 kW [Model: TTREX III, Rigaku, Japan]. The XRD spectra had also been collected in various temperatures (12≤T≤300 K) using the same powder diffractometer. The XRD patterns had been analyzed by the Lebail refinement method using the FULLPROF software [26]. Thermal transport [$\rho$(T), $S$(T), $\kappa$(T)], magnetic [M(T, H)] and heat capacity (in the absence of external magnetic field) measurements were performed in the temperature range 2 - 300 K using commercial set ups [Models: SQUID-VSM and PPMS Evercool-II, Quantum Design Inc., USA].

Electronic structure calculations were carried out using a full potential linearized augmented plane wave (FP-LAPW) method as implemented in WIEN2K package [27, 28]. Since the traditional exchange functional like LDA and GGA might underestimate the band gap, we have cross checked the band profile with the Tran-Blaha modified Becke-Johnson (TB- mBJ) functional [29, 30]. Spin-orbit coupling has been incorporated using a second variation scheme [31]. Transport coefficients such as Seebeck coefficient (thermopower) ($S$ in $\mu$V/K) and electrical conductivity scaled by relaxation time ($\sigma/\tau$ in $\Omega^{-1}m^{-1}s^{-1}$) were calculated using the BoltzTraP code [32] with a dense k- mesh of the order of 50×50×50 k-points. The BoltzTraP code is based on the rigid band approximation [33–35] and the constant scattering time approximation, and these approximations have been successfully applied earlier for several thermoelectric materials [36–40].

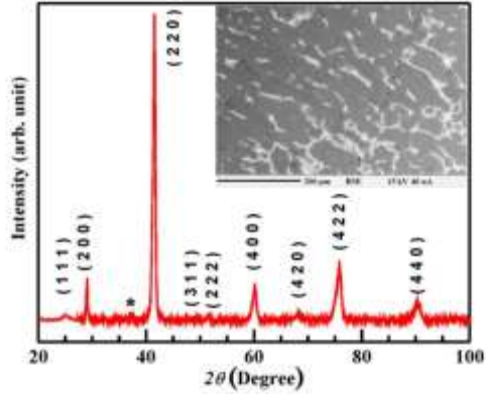

FIG. 1: Powdered X-ray diffraction pattern of Ru$_2$NbAl, measured at room temperature & indexed considering the L2$_1$ crystal structure, with a weak extra peak marked by an asterisk; Inset: Back scattered electron image of Ru$_2$NbAl.

## III. RESULTS AND DISCUSSION

### A. Structural details

The room temperature XRD pattern of Ru$_2$NbAl is presented in Fig. 1. Except a peak of negligible intensity (<2% of most intense peak) at ∼37°, all the other peak positions could be indexed by the L2$_1$ crystal structure (space group: $Fm\bar{3}m$) (Fig. 1) as suggested in the literature [23] and the lattice constant is found to be 6.1504(8) Å. The peak at ∼37° has however been found in many Ru based Heusler alloys and generally assigned to unreacted Ru [41, 42]. It has also been found that the presence of this minor phase in the material hardly influences their transport and magnetic properties [41, 42]. The EPMA measures a composition Ru$_{2.08(1)}$Nb$_{0.88(3)}$Al$_{1.04(3)}$ for the main Heusler phase and Ru$_{65}$Nb$_{33}$Al$_2$ for the other phase(s), distributed at the grain boundaries (Fig. 1, inset). Examination of the NbRu phase diagram [23] suggests that Ru$_{65}$Nb$_{33}$Al$_2$ is the spatially averaged composition of a eutectic which mixes the NbRu and Ru phases, in agreement with XRD.. The slightly nonstoichiometric composition of the main Heusler phase suggests the occurrence of Ru$_{Nb}$ and Al$_{Nb}$ antisite defects.

The low temperature XRD patterns taken in the range of 12 - 300 K do not show any significant change suggesting the invariance of crystal structure down to 12 K, the lowest temperature attainable in our diffractometer (Fig. 2). The lattice parameter gradually decreases as the temperature decreases. The unit-cell volume as a function of temperature [V(T)] is plotted and fitted (Fig. 2, inset) using the equation

$$V(T) = \gamma U(T)/K_0 + V_0, \quad (1)$$

where $V_0$ is the unit-cell volume at T = 0 K, $K_0$ represents the bulk modulus, and $\gamma$ is the Grüneisen parameter. U(T) is the internal elastic energy, generally expressed according to the Debye approximation as

$$U(T) = 9NK_BT\left(\frac{T}{\Theta_D}\right)^3 \int_0^{\Theta_D/T} \frac{x^3}{e^x - 1}dx \quad (2)$$

where N is the number of atoms per unit cell. Using this approximation, a Debye temperature $\Theta_D$ = 410 K and a Grüneisen parameter $\gamma$ = 1.8 have been estimated for Ru$_2$NbAl. This value of Debye temperature is consistent with the value further derived from heat capacity measurements ($\Theta_D$ = 418 K) (discussed later). This value is smaller than that, reported for Fe$_2$VAl ($\Theta_D$ = 540 K) [43] and most likely due to the heavier atomic mass of the chemical elements constituting Ru$_2$NbAl.

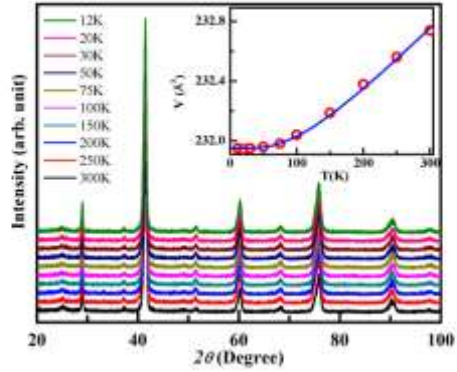

FIG. 2: Low temperature XRD pattern of Ru$_2$NbAl down to 12 K; Inset: Temperature dependence of unit-cell volume of Ru$_2$NbAl. Solid line represents a fit to Eq. 1.

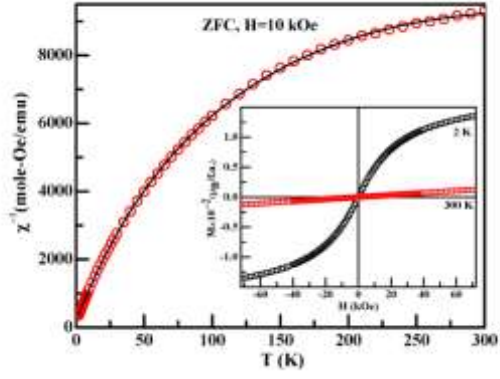

FIG. 3: Temperature dependence of inverse magnetic susceptibility of Ru$_2$NbAl measured in a 10 kOe applied magnetic field under ZFC configuration; Inset: Isothermal magnetization at 2 and 300 K of the same sample.

### B. Magnetic properties

#### 1. Magnetic susceptibility

To understand the magnetic properties of Ru$_2$NbAl, magnetic susceptibility ($\chi$) measurements have been carried out under both zero field cooled (ZFC) and field cooled (FC) configurations at H = 10 kOe and H = 70 kOe. The magnetic susceptibility exhibits no thermoremanence behavior between ZFC and FC protocol for both the magnetic fields. The absence of any anomaly in the magnetic susceptibility suggests the compound remains essentially paramagnetic down to the lowest measured temperature, 2 K (Fig. 3). The dominance of temperature independent Pauli paramagnetic behaviour is evident as the magnetization changes very slowly in the temperature range 300 - 100 K. On further lowering of the temperature below 100 K, magnetization starts to increase at a faster rate, particularly below 20 K. This behavior indicates the presence of a small localized paramagnetic contribution over a Pauli paramagnetic background.

To estimate this localized paramagnetic contribution, we have plotted the inverse susceptibility ($\chi^{-1}$) as a function of temperature in Fig. 3. In case of localized magnetic spins, $\chi^{-1}$(T) is known to exhibit a linear Curie-Weiss behavior. The inverse susceptibility of Ru$_2$NbAl could thus be very well fitted with a modified Curie-Weiss law

$$\chi(T) = \chi_0 + \alpha T^2 + \frac{C}{T - \theta_p}, \quad (3)$$

where $\chi_0$ represents temperature independent Pauli paramagnetic or diamagnetic contributions, while $\alpha T^2$ is the temperature dependent higher order contribution to Pauli paramagnetism, generally not considered in the zero order approximation [44]. The third term describes the standard Curie-Weiss expression. The fit of $\chi^{-1}(T)$ for H = 10 kOe yields $\chi_0$ = 6.76(4) ×10$^{-5}$ emu/mol-Oe, $\mu_{eff}$ = 0.27(2) $\mu_B$, $\theta_p$ = −1.61(7) K and $\alpha$ = 1.08(4) ×10$^{-10}$ emu/mol-Oe-T$^2$. The fitted parameters remain essentially the same for the H = 70 kOe measurement as well. Such small values of $\mu_{eff}$ and $\theta_p$ point towards the very weak nature of the localized spins in this compound.

### 2. *Isothermal magnetization*

The isothermal magnetization [M(H)] curve of Ru$_2$NbAl at 300 K is linear (Fig. 3, inset), as expected for a paramagnetic material. Interestingly, the M(H) curve at 2 K deviates from linearity and magnetization slowly approaches towards a saturation-like behavior at a field higher than 70 kOe but does not exhibit any hysteresis (Fig. 3, inset). We have also measured M(H) at 5, 10, 15 and 20 K in the field range 0 - 70 kOe (Fig. 4). As temperature increases, the non-linearity in the isothermal magnetization gradually gets weakened, becoming almost linear above 20 K. This is at variance with the absence of any signature of long range interactions down to 2 K as the $\chi(T)$ curve of Ru$_2$NbAl does not show any anomaly in this temperature range. This is also confirmed by the heat capacity measurement (discussed later) and the Arrott plot (M$^2$ *vs.* H/M) that fails to exhibit any spontaneous magnetization (Fig. 4, inset) even at 2 K. Generally, in systems that exhibit no long range ferromagnetic ordering, such S-shaped anhysteretic M(H) curve (at 2 K) could have its origin in short range ferromagnetic (FM) interactions or due to a superparamagnetic (SPM) state, or a combination of both [12, 45]. This additional magnetic interaction appears to develop only below 20 K, where the magnetic isotherms start to deviate from a linear behavior.

In order to investigate the origin of the saturation-like behavior observed at low temperatures, in our first attempts, the magnetic isotherms were analyzed considering short range FM interactions through the Néel-Brown (NB) [46] and the micromagnetic (MM) [47] models. However, none of the isotherms, not even at 2 K, could be analyzed with either the NB or MM model, forcing us instead to consider the presence of a SPM phase in this compound. We obtained reasonably good fit for all the M(H) curves, except that taken at 2 K, by considering the equation

$$M(H) = M_S L(x) + \chi H \qquad (4)$$

where $x = \mu H/k_B T$, $M_S$ represents the saturation magnetization, $\mu$ is the average magnetic moment per cluster, $L(x) = coth(x) - 1/x$ is the Langevin function and $\chi$ is the paramagnetic (PM) susceptibility [12, 45]. The first term in Eq. 4 denotes the magnetic behavior of the SPM component, while the second term arises from the paramagnetic phase present in this material. The results of these fits are listed in Table I. It can be seen from the table that for all the temperatures from 5 to 20 K, the magnetic moment of a SPM cluster is thus estimated to be ~4 $\mu_B$. The value of the saturation magnetization and the number of clusters per mole (N) are nearly 0.01 $\mu_B$/f.u. and $\sim 10^{21}$, respectively and remain closely constant throughout the temperature range 5 - 20 K. However, the magnetic isotherm at 2 K still does not yield a good fit and will be discussed later.

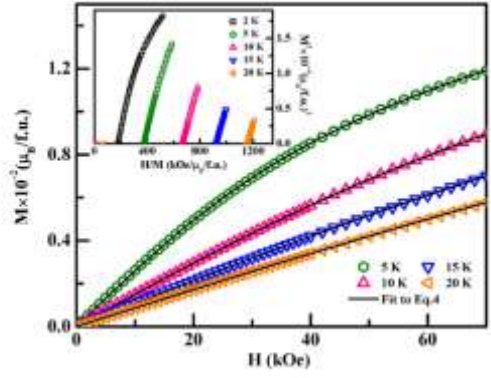

FIG. 4: Isothermal magnetization data at various temperature for $Ru_2NbAl$. The solid lines represent the fit of the data using Eq. 4; Inset: Arrott plots ($M^2$ vs. H/M) for $Ru_2NbAl$ at different temperatures.

To confirm the presence of a SPM state, a more rigorous check has been carried out by drawing a universal plot of reduced magnetization ($M/M_S$) as a function of H/T, where the magnetization at any particular temperature is normalized with respect to the saturation magnetization at the same temperature [12]. Since M(H) in $Ru_2NbAl$ exhibits both superparamagnetic as well as paramagnetic contributions, we have subtracted the estimated paramagnetic contribution from the experimentally obtained isothermal magnetization data [M(H)−$\chi$H] and normalized with respect to the saturation magnetization at the same temperature [(M(H)−$\chi$H)/M$_S$], which has been finally plotted as a function of H/T in the Fig. 5. All the isothermal curves up to 20 K, except that measured at 2 K, follow a single universal curve (Fig. 5). This results confirm the presence of a superparamagnetic state along with a paramagnetic state in the temperature range 5 - 20 K thus suggest the existance of non-interacting magnetic clusters in a PM matrix. The effect of inter-cluster interactions, if any, must be quite weak and masked by the SPM effect in this temperature region, 5≤T<20 K.

TABLE I: Parameters extracted from the fit of magnetic isotherms of Ru$_2$NbAl to Eq.4 for temperature range 5 - 20 K and SPM fit parameters from RA+SPM fit of 2 K data using Eq.7+M$_S$L(x).

| T (K) | $\mu$ ($\mu_B$) | M$_S$ ($\mu_B$/f.u.) | $\chi$ ($\mu_B$/f.u.) | M$_S/\mu$ (/f.u.) | N (/mol) |
|---|---|---|---|---|---|
| 2 | 3.883 | 0.00422 | 0 | 0.00108 | 0.65×10$^{21}$ |
| 5 | 4.348 | 0.01080 | 5.372×10$^{-4}$ | 0.00248 | 1.49×10$^{21}$ |
| 10 | 4.515 | 0.01032 | 4.570×10$^{-4}$ | 0.00228 | 1.37×10$^{21}$ |
| 15 | 4.721 | 0.00938 | 4.193×10$^{-4}$ | 0.00198 | 1.19×10$^{21}$ |
| 20 | 4.675 | 0.00920 | 3.767×10$^{-4}$ | 0.00196 | 1.18×10$^{21}$ |

As already discussed above, a deviation from Eq. 4 is observed in the M(H) curve measured at 2 K. The increase in both the magnetization value and the curvature of the magnetic isotherm at 2 K, can be attributed either to the blocking of SPM clusters or to the presence of interacting clusters that might have grown in strength below 5 K. Since there is no cusp observed in the ZFC curve down to 2 K (Fig. 3), the idea of the blocking of the SPM clusters can be ruled out, which indicates that the non-interacting SPM clusters start to interact or that some new interacting clusters develop at temperatures below 5 K. Existence of FM clusters at low temperature has also been reported in isostructural Heusler alloys, like Fe$_2$VAl, Fe$_2$V$_{1-x}$Cr$_x$Al [15, 17, 48]. In Fe$_2$VAl, based on density functional theory (DFT) calculations [18, 49], these ferromagnetic clusters were ascribed to Fe$_V$ and Fe$_{Al}$ antisite defects. To investigate the origin of interactions among clusters in our case, we have used the random anisotropy (RA) model [50], and analyzed the M(H) curve at 2 K. This model deals with the ground state configuration of magnetic materials having random anisotropy for a wide range of anisotropy strengths as well as experimentally applied magnetic fields. In case of weak anisotropy, three different regimes can be identified depending upon the relative strength of applied field (H). A parameter $H_s$ is used to gauge the relative strength of H and is expressed as

$$H_s = H_r^4/H_{ex}^3 \qquad (5)$$

where $H_r$ and $H_{ex}$ are the anisotropic field and exchange field, respectively. For low field region, H<$H_s$, one gets a correlated spin glass having large susceptibility. The intermediate-field regime, $H_s$<H<$H_{ex}$, is called the ferromagnet with wandering axis where the spins are nearly aligned. Any random anisotropy present in such system causes the directions of the magnetization of locally correlated regions to vary. In this regime, the magnetization approaches saturation as

$$M(H) = M_S^{FM}\left[1 - \frac{1}{15}\left(\frac{H_s}{H}\right)^{\frac{1}{2}}\right] \qquad (6)$$

where $M_S^{FM}$ is the saturation magnetization for ferromagnetic component. For high field region, H>$H_{ex}$, all spins are virtually aligned with the field, differing only by a small tipping angle that arises due to the random anisotropy. In such case, M(H) would gradually approaches towards $M_S^{FM}$ as

$$M(H) = M_S^{FM}\left[1 - \frac{1}{15}\left(\frac{H_r}{H + H_{ex}}\right)^2\right] \qquad (7)$$

The M(H) data at 2 K could be well fitted with Eq. 7 after considering an additional SPM contribution ($M_S L(x)$, $x = \mu H/k_B T$) (Fig. 5, inset). The fitted parameters obtained for the RA contribution are $M_S^{FM}$ = 0.0124 $\mu_B$, $H_r$ = 236.9 kOe and $H_{ex}$ = 61.0 kOe. The value of $H_r$ is larger than $H_{ex}$, which reveals that in our material the strength of the anisotropy is strong [50]. But surprisingly $H_r$ also exceeds the maximum applied field *i.e.*, H = 70 kOe. Here it may be noted that the usage of the prefactor 1/15 in Eq. 7, which was calculated for $H_r$<<$H_{ex}$ [50], can be applied only when the magnetization has reached over 93% of its saturation value [51]. However, in the present case, the saturation value of M(H) at 2 K and at 70 kOe field appears to be substantially lower. As a result, using the same prefactor of 1/15 is unlikely to be appropriate in our system that has much higher level of anisotropy and therefore requires a more suitable correction. A similar situation in case of $Dy_xY_{1-x}Al_2$ [51], $GdAl_2$ [52] and $Co_{58.5}Ga_{41.5}$ [53] compounds had earlier been dealt by considering the prefactor to be of the order of 1. Accordingly, by considering a prefactor of order unity, $H_r$ reduces to 61.2 kOe, although interestingly $H_{ex}$ remains the same. The reduced value of $H_r$ and $H_{ex}$ are less than the maximum applied field (70 kOe) and thus fulfills the condition for materials having strong anisotropy as demanded by Eq. 7. The presence of such strong anisotropy would lead to make the system as speromagnetic-like, where the local magnetizations generally follow the local anisotropy axis [50]. However, in our case, the occurrence of speromagnetic-like state is forbidden as $H_r \leq H_{ex}$. Rather the system is more likely to have ferromagnetically correlated regions (clusters), whose magnetization directions are pinned or frozen by random anisotropy, as in a correlated glassy system.

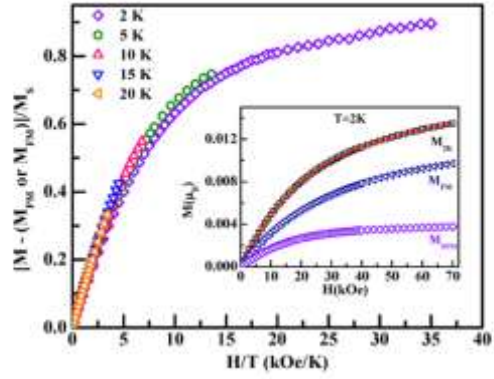

FIG. 5: The SPM state follows universal curve at low temperature isotherm for $Ru_2NbAl$; Inset: Isothermal magnetization data at 2 K along with a fit (solid line) to the Eq. 7 and the individual FM and SPM contributions to the same isotherm.

As mentioned above, in addition to the contribution from the correlated magnetic clusters, the M(H) data at 2 K also contains an additional contribution from SPM clusters (Fig. 5, Inset). The SPM component obtained by subtracting the RA part, when normalized by $M_S$ and plotted against H/T, follows the same universal curve discussed earlier (Fig. 5). The resulting parameters correspond to SPM state at 2 K are also tabulated in Table I. Although the magnetic moment per cluster remains near-about same, the value of $M_S$ and number of SPM clusters are reduced significantly. The reduction of these superparamagnetically interacting clusters affirms that FM interaction develops among the rest of the clusters. The above analysis suggests that in case of Ru$_2$NbAl, although the SPM component exists down to 2 K whereas, FM interactions becomes prevalent below 5 K. Above 20 K, the magnetization is dominated by the Pauli paramagnetic component.

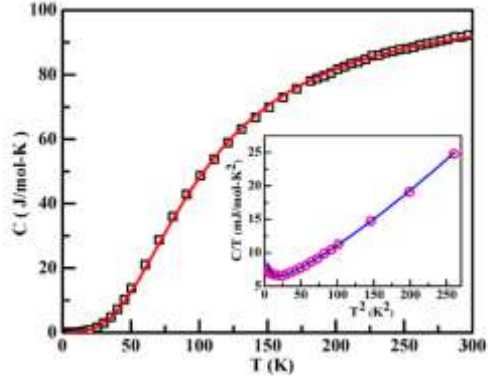

FIG. 6: Specific heat as a function of temperature of Ru$_2$NbAl. Solid line represents the fit to Eq. 8; Inset: C/T vs T$^2$ plot at low temperature along with a fit to Eq. 10.

### C. Heat capacity

In order to verify the presence of magnetic clusters in Ru$_2$NbAl using another independent experimental probe, we have carried out heat capacity [C(T)] measurement in the temperature range 2 - 300 K in absence of any external magnetic field (Fig. 6). No peak could be observed in the C(T) curve (Fig. 6) in the measured temperature range, which is also an indication of the absence of any long range magnetic order, similar to that inferred from magnetic susceptibility measurements.

Generally, the heat capacity of a metallic system can be written as

$$C(T) = \gamma T + 9nR\left(\frac{T}{\Theta_D}\right)^3 \int_0^{\frac{\Theta_D}{T}} \frac{x^4 e^x}{(e^x-1)^2} dx \qquad (8)$$

where the first term represents the electronic specific heat and the second term comes from the lattice/phonon contribution. $\gamma = \frac{1}{3}\pi^2 D(E_F) k_B^2$, is the Sommerfeld coefficient, where D(E$_F$) is the density of states at the Fermi level E$_F$. $n$ is the number of atoms per formula unit (for Ru$_2$NbAl: $n = 4$), $\Theta_D$ is the Debye temperature and $x = \hbar\omega/k_B T$. The standard Debye model, discussed above, could explain the heat capacity data very well in the temperature range 25 - 300 K, by considering $\gamma$ = 3.7 mJ/mol-K$^2$ and $\Theta_D$ = 418 K, in agreement with the value derived from the lattice parameter measurements as a function of temperature. The low temperature data, however, deviates from the standard behavior.

In the low temperature region Eq. 8 can be simplified as

$$C(T) = \gamma T + \beta T^3 + \delta T^5 + \cdots \cdots \quad (9)$$

where $\beta$, $\delta$ are the coefficients. Below $\sim \Theta_D/50$, $\delta T^5$ and other higher order terms could be neglected [54] and the heat capacity behavior in the representation of C/T vs T$^2$ is expected to show a linear dependence: this is indeed the case above 5K (Fig. 6, inset). However, an upturn observed in C/T vs T$^2$ (Fig. 6, inset) plot of Ru$_2$NbAl below 5 K, suggests the presence of additional contribution to the heat capacity given in Eq. 9. It may be noted here that in Ru$_2$NbAl, the development of ferromagnetically interacting clusters below 5 K was inferred from the isothermal magnetic measurements. Therefore, the additional contribution to heat capacity in Ru$_2$NbAl appears to have originated from the inter/intra clusters interactions.

Such upturn in the low temperature heat capacity has also been found in Fe$_2$VAl [55], Fe-V [56], TiFe alloys [57] where oscillation of small FM clusters have been argued to be responsible for this behavior. In those compounds, it was claimed that the system having ferromagnetic clusters in nonferromagnetic matrix normally rests in a position of minimum energy and the potential energy of the system gets enhanced when the direction of the magnetization vector of the clusters are altered due to the application of any force. The enhanced energy is generally stored in the system through the local elastic deformation of the clusters and the matrix as well as in magnetostriction energy [57]. So, due to thermal excitation, each cluster makes oscillation about a direction determined by its crystallographic anisotropy energy and absorbs $k_B$T amount of thermal energy. This gives rise to an extra constant term $C_0 \sim 2k_BN$ to the total specific heat [55, 56], where $N$ is the number of such oscillating magnetic clusters. However, to satisfy the third law of thermodynamics, which requires the heat capacity to be zero at T = 0 K, it is argued that this additional contribution, $C_0$, gradually loses its strength below a certain temperature, called the Einstein temperature, $T_E = 2\beta H/k_B$, where H is the magnetic field needed to produce the same torque as the crystal anisotropy energy [56]. $T_E$ is generally found to be very low, <1 K in most of the materials [56, 58]. So, by considering the cluster interaction term, Eq. 9 gets modified to

$$C(T) = \gamma T + \beta T^3 + \delta T^5 + C_0 + \cdots\cdots \quad (10)$$

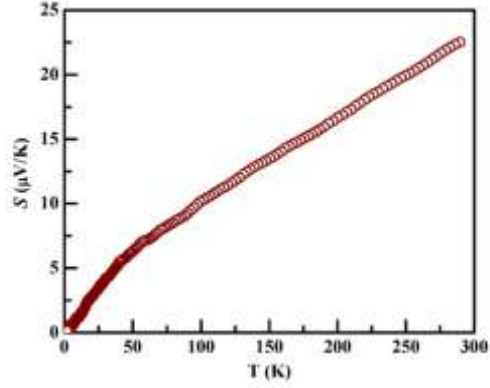

FIG. 7: Temperature dependance of Seebeck coefficient of $Ru_2NbAl$ in the absence of magnetic field. Application of magnetic field of 80 kOe (not shown in the figure) does not have noticeable influence in the $S$ (T) behaviour.

This modified equation could reproduce quite well the experimentally observed specific heat data of $Ru_2NbAl$ in the temperature range 2 - 20 K (Fig. 6, inset). We have included the $\delta T^5$ term in the fit so as to cover the extended temperature range beyond $\Theta_D/50$. The estimated values of the parameters from this fit are $\gamma$ = 2.7 mJ/mol-$K^2$, $\beta$ = 0.067 mJ/mol-$K^4$, $\delta$ = 6×$10^{-5}$ mJ/mol-$K^6$ and $C_0$ = 11.04 mJ/mol-K. Using the value of $C_0$ we have calculated the number of ferromagnetic clusters as $N$ ~4×$10^{20}$ per mol. This number matches quite closely to the number of FM clusters (~8×$10^{20}$) estimated earlier from the magnetization data at 2 K (Table I).

### D. Transport properties

*1. Seebeck coefficient*

To study the thermoelectric properties of $Ru_2NbAl$, we have carried out Seebeck coefficient (thermopower), resistivity and thermal conductivity measurements in the absence (H

= 0) and presence of a 80 kOe magnetic field. The magnetic field is found to have very insignificant impact on all the three measurements. The temperature dependent Seebeck coefficient $S(T)$ of Ru$_2$NbAl at H = 0 is only shown in Fig. 7. The measured value of $S$ remains positive in the entire temperature range under investigation suggesting that the dominant carriers in thermoelectric transport are holes in this compound. The Seebeck coefficient at 300 K is equal to 22 $\mu$V/K, a value comparable to those reported for other Heusler alloys like Fe$_2$VAl (~35 $\mu$V/K), Fe$_2$VGa (~30 $\mu$V/K) and Ru$_2$NbGa (~20 $\mu$V/K) [59–64]. Such a moderate value at room temperature, combined with the linear increase from ~50 K to 300 K is a characteristic of metallic state and suggests the possibility of a semi-metallic ground state for Ru$_2$NbAl, in agreement with dominant Pauli paramagnetism above 20 K and the non-zero value of Sommerfeld coefficient $\gamma$ = 2.7 mJ/mol-K$^2$. Moreover, as will be further reported below, $S_{300K}$ = 22 $\mu$V/K is also consistent with the value obtained by DFT calculations which also conclude that Ru$_2$NbAl is a semi-metal.

*2. Resistivity*

The electrical resistivity as a function of temperature [$\rho$(T)] of Ru$_2$NbAl, studied in the temperature range 2 - 300 K at H = 0 is shown in Fig. 8. There is no significant change in the data in presence and absence of a magnetic field, not even exhibiting any thermal hysteresis behavior. The value of resistivity is 452 $\mu\Omega$-cm at room temperature, whereas it is found to be ~538 $\mu\Omega$-cm at 2 K, indicating that this quantity varies within a narrow range of values. These values are rather characteristic of Ru$_2$NbAl being a semi-metal or a degenerate semi-conductor, but surprisingly, the $\rho$(T) curve exhibits a negative temperature coefficient of resistivity (TCR), characteristic of a semiconductor. Such activated behaviour for the resistivity has also been reported in other Heusler alloys having VEC 24 *viz.*, Fe$_2$VAl [8], Ru$_2$NbGa [42], Ru$_2$TaAl [65] *etc*. A plot of ln($\rho$) *vs*. 1/T reveals that the linear region is only found in the temperature range 225 - 300 K (Fig. 8, inset(a)). The value of activation energy ($\Delta$), estimated from the slope of the curve as $\Delta$ ~5.3 meV ~63 K. This small value cannot be ascribed to the intrinsic band gap which would otherwise manifests itself by a non-monotonous variation of the Seebeck coefficient at low temperature (minority carrier effect).

The resistivity data at lower temperature (T<225 K) increases at a much slower rate than that expected (Fig. 8, dashed line) for an activated behaviour. Such deviation may be attributed to the temperature dependence of carrier mobility [20]. Alternatively, negative TCR at low temperature can also be explained by variable range hopping (VRH) conduction proposed by Mott [66]. In this mechanism electrons hop to energetically closed and localized states and conduction law of VRH can be expressed as

$$\rho(T) = \rho_0 + A exp\left[\left(\frac{T_0}{T}\right)^{1/4}\right] \qquad (11)$$

where $A$ is a constant and $T_0$ is the activation temperature that depends on the localization length ($\xi$) as $\xi^{-3}$ [66]. Using the above equation, the resistivity data of Ru$_2$NbAl can be fitted below 60 K (Fig. 8, inset(b)). The analysis yields $T_0$ to be 0.25 K, a similar value (0.063 K) of the activation temperature was earlier found in Fe$_2$V$_{1-x}$Nb$_x$Al [67]. These results are rather suggestive of localization effects of the charge carriers by structural disorder rather than of a true semi-conducting ground state, as already discussed in the literature on Fe$_2$VAl [17, 48]. The exact scenario of such localization remains currently elusive but both EPMA and magnetic measurements indicate the occurrence of structural defects in Ru$_2$NbAl. The latter technique shows the existence of $\sim 10^{21}$ superparamagnetic (SPM) clusters. Since in Fe$_2$VAl, the magnetic defects are antisite Fe$_V$ and Fe$_{Al}$ [18], by analogy, Ru$_{Nb}$ and Ru$_{Al}$ antisite defects in Ru$_2$NbAl could also be magnetic and give rise to the SPM clusters detected by magnetization measurements. EPMA indeed indicates the occurrence of Ru$_{Nb}$ and Al$_{Nb}$ antisite defects which could play a role in the semiconducting-like behavior of the electrical resistivity.

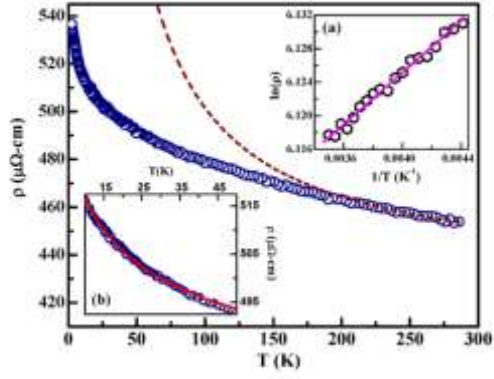

FIG. 8: The temperature dependent resistivity behaviour for Ru$_2$NbAl at H = 0; Inset (a): ln($\rho$) versus 1/T plot at temperature range 225 - 300 K. Solid line depicts a linear fit; Inset (b): Resistivity data below 60 K with a fit to Eq. 11.

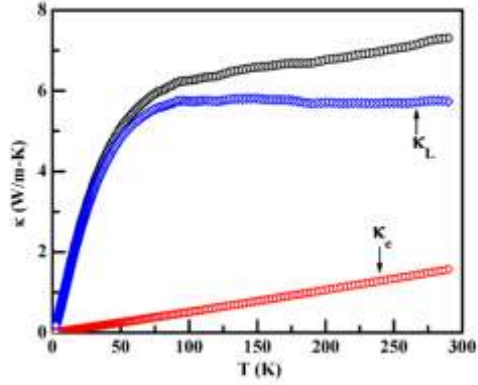

FIG. 9: Temperature variations of the total thermal conductivity $\kappa$, lattice thermal conductivity $\kappa_L$, and electronic thermal conductivity $\kappa_e$ for Ru$_2$NbAl at H = 0.

### 3. Thermal conductivity

To further evaluate the thermoelectric performance of Ru$_2$NbAl, thermal conductivity ($\kappa$) is measured between 2 - 300 K, shown in Fig. 9. At low temperatures, $\kappa$ increases rapidly with temperature which is typical for solids as thermal scattering by mass defects (isotopes, impurity, *etc.*) increases with temperature [68]. Above 50 K, the rate of increment has been slowed down and the value of $\kappa$ reaches 7.3 W/m-K at 300 K (Fig. 9). In general, the total thermal conductivity of metals and semi-metals is defined by a sum of electronic ($\kappa_e$) and lattice ($\kappa_L$) contributions. The electronic thermal conductivity can be estimated using the Wiedemann-Franz law $\kappa_e \rho/T = L_0$, where $\rho$ is the measured electric resistivity and $L_0 = 2.45 \times 10^{-8}$ W$\Omega$/K$^2$ is the Lorenz number. The lattice thermal conductivity, $\kappa_L$, thus can be evaluated by subtracting $\kappa_e$ from the observed $\kappa$. The value of $\kappa_e$ thus found to be very small, suggesting that the thermal conductivity of this compound is necessarily due to $\kappa_L$ (Fig. 9). At low temperatures $\kappa_L$ increases with temperature and a maximum appears between 50 K and 100 K due to the reduction in thermal scattering at low temperatures. The $\kappa_L$ value of Ru$_2$NbAl at 300 K is estimated to be 5.6 W/m-K (Fig. 9), which is much lower than those found in Fe$_2$VAl ($\sim$28 W/m-K) and Fe$_2$VGa ($\sim$17 W/m-K) [61–64]. This reduction may be attributed to phonon scattering caused by the elemental substitution by heavier Ru and Nb atoms in the place of Fe and V in Fe$_2$VAl [20, 65]. Antisite defect arising from the elemental substitutions is strongly related to crystal lattice strain and may also be responsible for the reduction of $\kappa_L$ in Ru$_2$NbAl. This low value of $\kappa$ observed in Ru$_2$NbAl, makes it a potential candidate for thermoelectric applications.

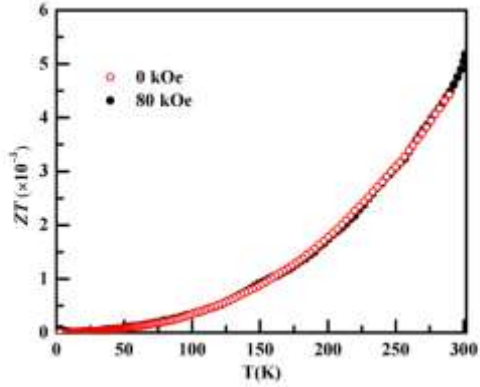

FIG. 10: *ZT* value as a function of temperature for $Ru_2NbAl$ in the absence (H = 0) and presence (H = 80 kOe) of magnetic field.

### 4. *Figure of merit*

Since $Ru_2NbAl$ exhibits a smaller thermal conductivity and a comparable power factor ($S^2/\rho$) to other full Heusler alloys, its dimensionless figure of merit $ZT = 5.2 \times 10^{-3}$ (Fig. 10) is larger at room temperature than that in those alloys. Nonetheless, it is still several orders of magnitude smaller than that of the state-of-the-art thermoelectric material at 300 K, $Bi_2Te_3$ which displays $ZT = 1$. Doping, to adjust the charge carrier concentration and maximize the power factor and substitution, to further decrease the thermal conductivity, will be required to improve *ZT* in $Ru_2NbAl$.

### E. Electronic Structure Calculations

The experimental structure parameters were optimized, and further calculations are per- formed using those optimized parameters. We have performed the electronic structure calculations using several exchange functional like LDA, GGA and TB-mBJ, and the calculated band structure using TB-mBJ functional is presented in Fig. 11. Spin orbit coupling has been included in the calculations due to the presence of heavy elements. From the band structure it is evident that the compound possesses semi-metallic character, as earlier argued from the experimental observations. The bands cross the Fermi level around $\Gamma$ and X high symmetry points. The conduction and valence bands are found to touch around the X point, and preserve a gap in all other points in the Brillouin zone. Fig. 12(a) shows the density of states (DOS) of the investigated compound, together with the partial density of states. Highly competing Ru and Nb "$d$" states are found near the Fermi level, as shown in the figure (Fig. 12(a)). A total of four bands are crossing the Fermi level, and among these bands three are found to be of hole-like nature and the other one found to possess electron-like nature. One may note here that our Seebeck coefficient measurements have also suggested that the majority charge careers are holes. The merged Fermi surface plot of the investigated compound is represented in Fig. 12(b). As mentioned before, there are pockets in the Fermi surface around $\Gamma$ and X points. The density of states variation above and below Fermi level are almost similar, indicating the possibility of both hole and electron carriers for thermoelectric applications, which certainly attracts device applications, but it is to be mentioned that the DOS is steeply rising for hole carriers and we base our discussion only on holes in this work.

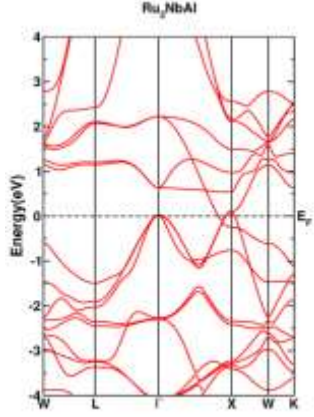

FIG. 11: Calculated band structure using TB-mBJ functional.

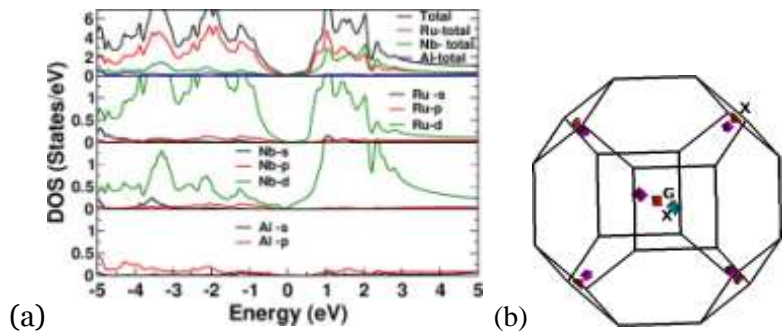

FIG. 12: Calculated (a) density of states (b) Merged Fermi surface.

The mechanical properties are examined, and the calculated parameters are listed in Table II. The high value of bulk modulus indicate the stiffness of the compound. The calculated Debye temperature is found to be little higher than the experimental value.

TABLE II: Calculated mechanical properties.

| | |
|---|---|
| $C_{11}$(GPa) | 405.86 |
| $C_{12}$(GPa) | 152.81 |
| $C_{44}$(GPa) | 96.84 |
| Bulk modulus(GPa) | 237.40 |
| Debye temperature(K) | 535.917 |

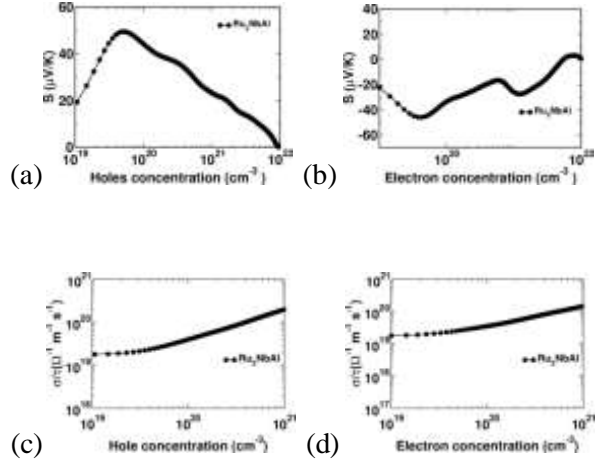

FIG. 13: Variation of Seebeck coefficient ((a) & (b)) and electrical conductivity ((c) & (d)) as a function of hole and electron concentrations.

Thermoelectric coefficients like Seebeck coefficient, electrical conductivity scaled by relaxation time and power factor are calculated by combining semi-classical Boltzmann transport equation with density functional theory. Calculated Seebeck coefficient as a functions of holes and electrons are given in Fig. 13(a)&(b). For both holes and electrons, carrier concentrations around $5\times10^{19}$ cm$^{-3}$ secure the maximum value of Seebeck coefficient. The temperature dependent Seebeck coefficient for hole concentration around $1\times10^{18}$cm$^{-3}$ is rep- resented in Fig. 14. The trend is in agreement with the experiment. In addition, we have deduced the electrical conductivity scaled by relaxation time, and the carrier concentration dependency of the same is also represented in Fig. 13(c)&(d). Using the calculated Seebeck coefficient and electrical conductivity scaled by relaxation time, we have analyzed the power- factor value. The experimental resistivity value has been adapted to find the conductivity value, and we have decoupled the relaxation time. The estimated relaxation time turned out to be around $1\times10^{-14}$ s. The calculated figure of merit around

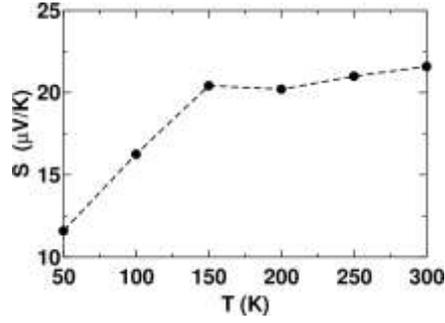

FIG. 14: Calculated temperature dependent Seebeck coefficient value.

300 K is $4.29\times10^{-3}$, which is in good agreement with the experimental value.

## IV. CONCLUSION

The magnetic properties of Ru$_2$NbAl are dominated by intrinsic Pauli paramagnetism and superparamagnetism of magnetic defects below $\sim$20 K. These magnetic defects as well as the conduction electrons ($\gamma$ = 2.7 mJ/mol-K$^2$) are found to have discernible contribution to the specific heat at low temperature. The Seebeck coefficient displays moderate positive values and linearly increases with temperature. All these experimental results suggest a semi-metallic ground state for Ru$_2$NbAl, in agreement with the DFT calculations. In this context, the activated behavior of the electrical resistivity would thus arise from charge carrier localization due to structural antisite defects, detected by both magnetic measurements and EPMA. Ru$_2$NbAl displays a thermal conductivity smaller than that found in other full Heusler alloys and could, after suitable doping, display a larger thermoelectric figure of merit.

## V. ACKNOWLEDGEMENT

The authors SPC and VK would like to thank IIT Hyderabad for computational facility, and SPC would like to thank MHRD, Govt. of India for fellowship.

---